\documentclass[12pt,preprint]{aastex}

\slugcomment{Submitted on: \today }

\shorttitle{The Effect of Lithium Recombination on CMB Anisotropies}
\shortauthors{Stancil et al.}

\begin{document}

\title{Cosmological Recombination of Lithium and its Effect on the
Microwave Background Anisotropies}

\author{Phillip C. Stancil\altaffilmark{1},
       Abraham Loeb\altaffilmark{2}, 
       Matias Zaldarriaga\altaffilmark{3}, Alexander
       Dalgarno\altaffilmark{2}, \\ and Stephen Lepp\altaffilmark{4} }

\altaffiltext{1}{Department of Physics and Astronomy and Center
       for Simulational Physics, The University of
       Georgia, Athens, GA 30602-2451; stancil@physast.uga.edu}

\altaffiltext{2}{Harvard-Smithsonian Center for Astrophysics,
                 60 Garden St., Cambridge, MA 02138; 
                 aloeb@cfa.harvard.edu, adalgarno@cfa.harvard.edu} 

\altaffiltext{3}{Physics Department, New York University, 4 Washington
                 Place, New York, NY 10003; matiasz@physics.nyu.edu}

\altaffiltext{4}{Department of Physics, University of Nevada, 
                 Las Vegas, NV 89154-4002; lepp@nevada.edu}

\begin{abstract}

The cosmological recombination history of lithium, produced during Big--Bang
nucleosynthesis, is presented using updated chemistry and cosmological
parameters consistent with recent cosmic microwave background (CMB)
measurements.  For the popular set of cosmological parameters, about a
fifth of the lithium ions recombine into neutral atoms by a redshift $z\sim
400$.  The neutral lithium atoms scatter resonantly the CMB at 6708~\AA~ and
distort its intensity and polarization anisotropies at observed wavelengths
around $\sim 300~\mu$m, as originally suggested by Loeb (2001).  The
modified anistropies resulting from the lithium recombination history are
calculated for a variety of cosmological models and found to result
primarily in a suppression of the power spectrum amplitude. Significant
modification of the power spectrum occurs for models which assume a large
primordial abundance of lithium. While detection of the lithium signal
might prove difficult, if offers the possibility of inferring the lithium
primordial abundance and is the only probe proposed to date of the
large-scale structure of the Universe for $z\sim 500-100$.

\end{abstract}


\section{Introduction}

Lithium is the heaviest stable element produced during Big--Bang
nucleosynthesis \citep*[and references therein]{bur01}. However, due to its
low primordial abundance relative to hydrogen $X_{\rm Li}\sim
10^{-10}$--$10^{-9}$, its significance in the early Universe was thought to
be restricted only to the formation of rare molecules such as LiH
\citep*[and references therein]{sta96}. Hence, the recombination history of
lithium itself was only calculated as an intermediate step in a chain of
chemical reactions and under a set of simplifying assumptions
\citep*[hereafter SLD96, SLD98]{pal95,sta96,sta98}. These preliminary
calculations indicated that a substantial fraction ($\ga 20$\%) of the
singly-charged lithium ions formed neutral atoms by recombination 
in the redshift interval $z\sim 400$--$500$.

Recently, \citet{loe01} has shown that the formation of neutral lithium can
strongly modify the anisotropy maps of the cosmic microwave background
(CMB) through the absorption and re-emission at its resonant
6708~\AA~transition from the ground state.  Despite the exceedingly low
lithium abundance\footnote{Note that by the redshift of interest, all the
$^7$Be produced during Big--Bang nucleosynthesis has been converted to
$^7$Li.}  left over from the Big Bang, the resonant optical depth after
lithium recombination is expected to be as high as $\tau_{_{\rm LiI}}\sim
0.4 (X_{\rm Li}/3.8\times 10^{-10})$ at a redshift $z\sim 400$ if half of
the lithium ions recombine by then. The scattering refers to an observed
wavelength of $\lambda(z)=[6708~$\AA$\times (1+z)]=268.3~\mu{\rm m}\times
[(1+z)/400]$, where $X_{_{\rm Li}}\approx 3.8\times 10^{-10}$ is the latest
estimate of the lithium to hydrogen number density ratio \citep{bur01}.
\citet{loe01} argued that resonant scattering would suppress the original
anisotropies by a factor of $\exp({-\tau_{_{\rm LiI}}})$, but will generate
new anisotropies in the CMB temperature and polarization on sub--degree
scales primarily through the Doppler effect.  Observations at different
far--infrared wavelengths could then probe different thin slices of the
early Universe. \citet{zal02} calculated in detail the expected
anisotropies in both the temperature and polarization of the CMB and
assessed their detectability relative to the far-infrared background. They
concluded that the modified polarization signal could be comparable to the
expected polarization anisotropies of the far-infrared background on
sub-degree angular scales ($\ell \ga 100$).  However, these calculations
assumed a range of trial values for the neutral fraction of lithium in the
redshift range $z=400$--500 lacking knowledge of its full redshift
dependence.

In this paper, we calculate rigorously the recombination history of lithium
and its subsequent effect on the CMB anistoropies. In \S 2 we discuss the
physics responsible for the recombination history of lithium and in \S 3 we
obtain the resulting optical depth for the 6708~\AA~transition.  \S 4
describes the effects of the lithium optical depth on the CMB power spectra.
Finally, \S 5 summarizes this work.

\section{The Lithium Recombination History}

In order to assess the importance of the optical depth of lithium on the
CMB anisotropies, an accurate determination of the neutral lithium
abundance as a function of redshift is required.  The post--recombination
abundance of Li was first addressed by \citet{lep84} in their investigation
of molecules formed in the early Universe. While this work gave no
information on the Li and Li$^+$ abundances or the adopted rate
coefficients, they were implicitly needed to obtain the LiH abundance, a
main focus of their study. \citet{puy93} also investigated the LiH
post-recombination abundance, but neglected to give any details concerning
the evolution of Li and Li$^+$.

The first explicit results for the redshift dependent Li and Li$^+$
abundances were presented by \citet{pal95} although no information was
given on the adopted chemistry or reaction rate coefficients.  Their
results, which varied little with the adopted cosmological model
parameters, suggested that the neutral Li formation redshift was $z_{\rm
f}\sim 500$ (which we define here as the redshift when the neutral
abundance reaches 10\% of the elemental abundance $n_{\rm Li}$) and that
the Li ionization fraction $n$(Li$^+$)/$n_{\rm Li} \to 0.16$ as $z\to 0$.
The significant residual Li$^+$ abundance, a result predicted by
\citet{dal87}, is a consequence of the depletion
of electrons following the cosmological recombination of 
hydrogen and helium.  The
related chemistry was described in more detail and improved upon
in SLD96, \citet{bou97}, and
\citet[hereafter GP98]{gal98}.  The primary Li formation and destruction
mechanisms are radiative recombination
\begin{equation}
{\rm Li}^+ + {\rm e}^- \rightarrow {\rm Li} + \nu,
\label{eq1}
\end{equation}
(L5)\footnote{The reaction labels (L$x$) correspond
to the process $x$ in SLD96.} and photoionization,
\begin{equation}
{\rm Li} + \nu \rightarrow {\rm Li}^+ + {\rm e}^-,
\label{eq2}
\end{equation}
(L10) respectively. For the former, they adopted the rate coefficients of
\citet[VF96]{ver96b} which is based on R-matrix calculations of the
photoionization cross sections of Li \citep*{pea88},
but adjusted to match experiment, and smoothly fit to higher-energy
Hartree-Dirac-Slater calculations \citep{ver96a}. Through comparison
of their photoionization cross sections to experiment, VF96 suggest that
their Case A recombination coefficients are accurate to better than
10\%. The Li
photoionization rate for a black-body radiation field was then obtained by
detailed balance from the radiative recombination rate coefficients (see
GP98). Using the same cosmological models of \citet{pal95} (see also Model
III of SLD96), \citet{bou97} found similar results, $z_{\rm f}\sim 500$ and
$n$(Li$^+$)/$n_{\rm Li}= 0.18$ at $z=10$. However, with further
improvements in the chemistry, GP98 predicted $z_{\rm f}\sim 400$ and
$n$(Li$^+$)/$n_{\rm Li}= 0.56$ at $z=10$.

\begin{figure}
\plotone{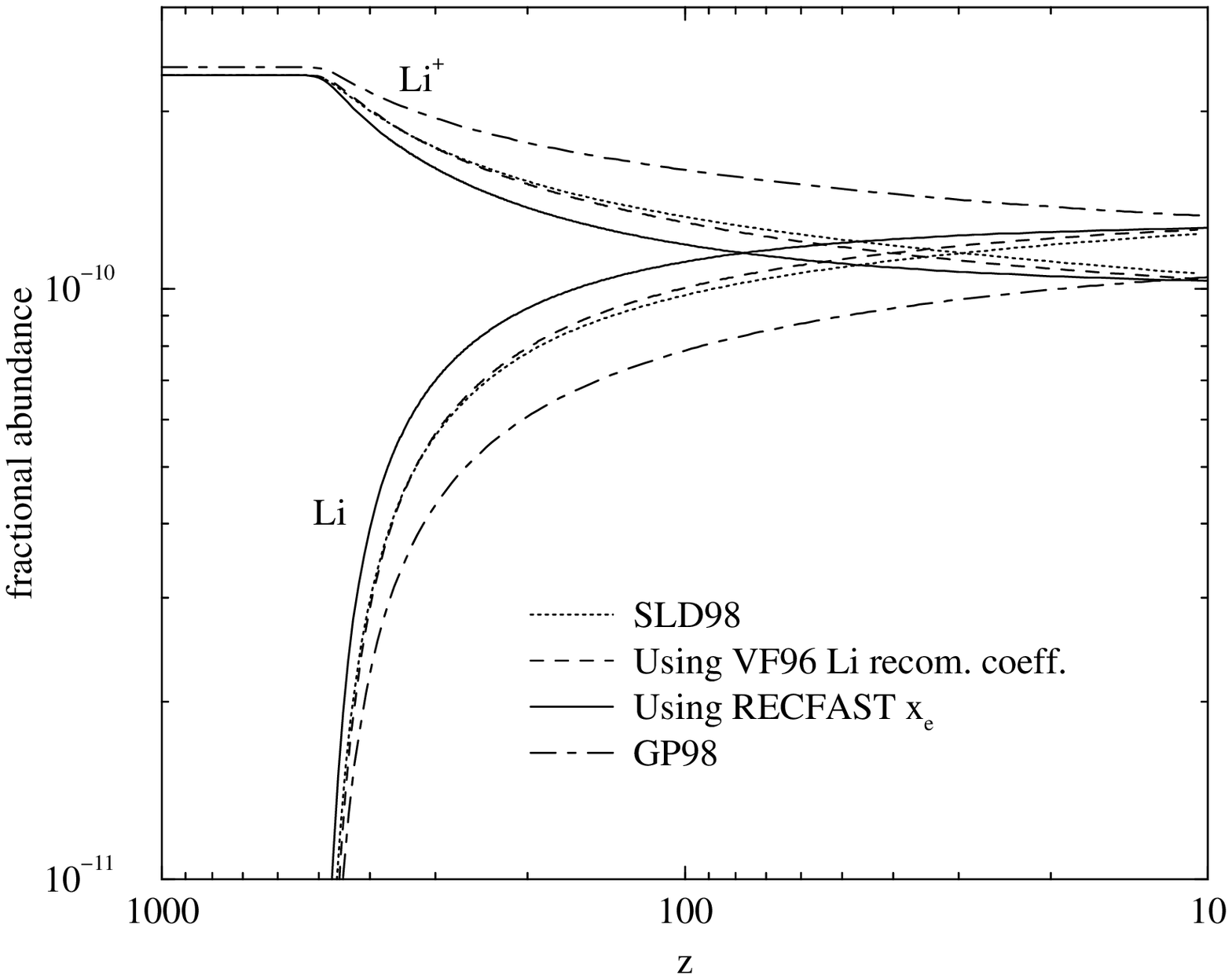}
\caption{Comparsion of various calculations of the Li and Li$^+$ fractional
abundance (relative to hydrogen) for the cosmological Model III of SLD96
($\Omega_0=1$, $\Omega_{\rm b}=0.0367$, $h=0.67$, $\Omega_{\rm b}h^2=0.0165$,
$Y=0.242$, and $X_{\rm Li}=2.3\times 10^{-10}$).
\label{fig1}}
\end{figure}

In their comprehensive chemistry 
SLD96 adopted the radiative recombination
coefficients of \citet{cav72} determined from their
model potential calculations of the photoionization
cross sections. SLD96 also estimated the
photoionization rate by detailed balance and
found $z_{\rm f}\sim 450$ and
$n$(Li$^+$)/$n_{\rm Li}= 0.34$ at $z=10$
(using the same cosmological parameters, Model III).
Following improvements in the lithium chemistry
\citep[see][]{sta97,sta98a,sta96a}, SLD98
obtained $z_{\rm f}\sim 440$ and
$n$(Li$^+$)/$n_{\rm Li}= 0.48$ at $z=10$.
Figure~\ref{fig1} displays the lithium abundances 
as a function of redshift from various calculations
using Model III of SLD96. 

\begin{figure}
\plotone{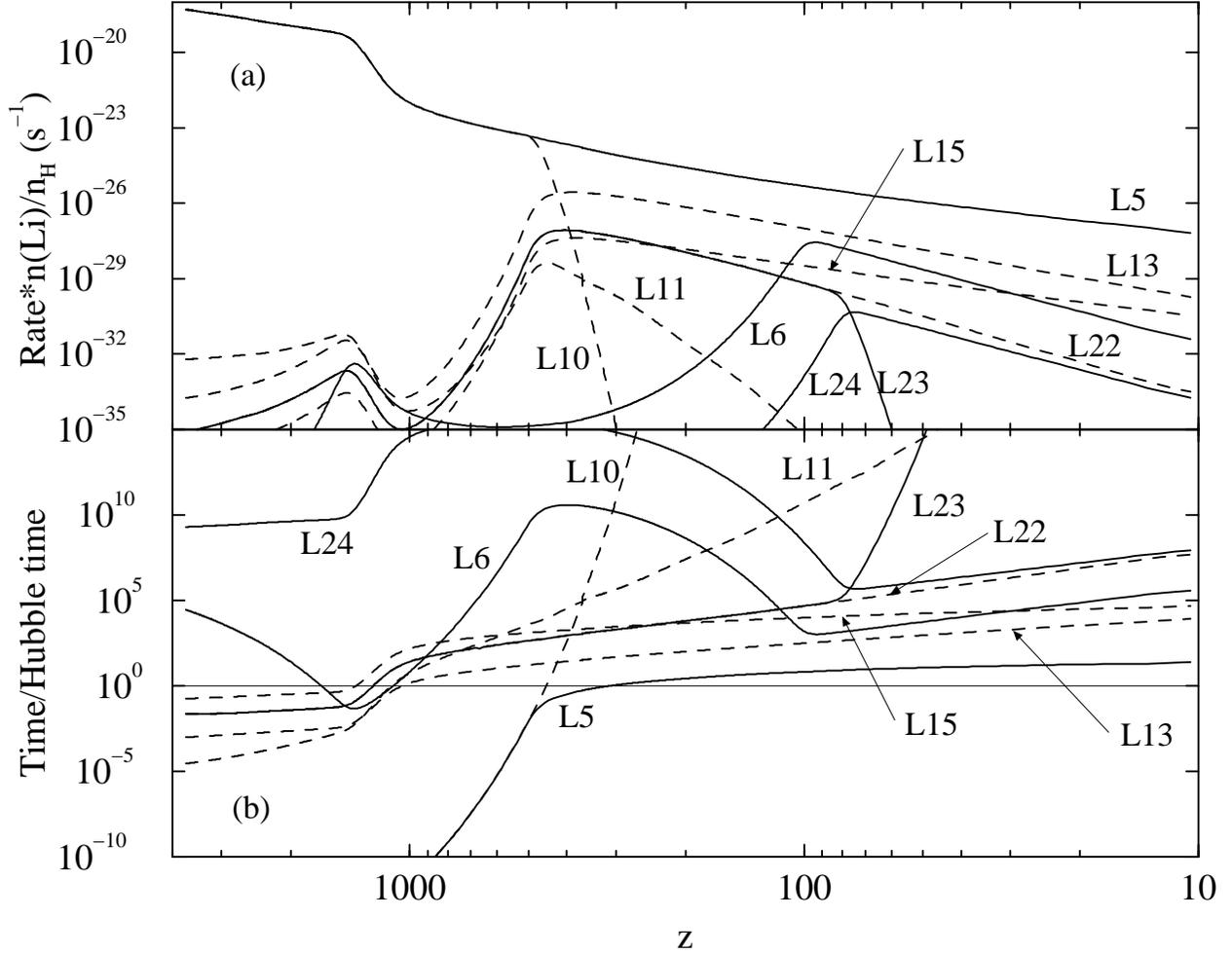}
\caption{Dominant formation (solid lines) and destruction (dashed lines)
processes for neutral lithium for Model III of SLD96. Panel (a) shows the
rates (multiplied by the neutral Li fractional abundance $n({\rm
Li})/n_{\rm H}$) and (b) shows the time scale per Hubble time $1/H(z)$ per
Li atom.  The reaction labels (L$x$) correspond to the process $x$ listed
in SLD96.
\label{fig2}}
\end{figure}

To explore the dependencies of the models on the adopted rate coefficients,
we repeated the calculations of SLD98, but using the rate coefficients for
reactions~(\ref{eq1}) and (\ref{eq2}) from GP98. The small changes shown in
Figure~\ref{fig1} reflect the fact that the radiative recombination rate
coefficients of VF96 are only about 5\% larger than those determined by
\citet{cav72}, which is however within the 10\% accuracy claimed by
VF96. Any differences in the rate coefficients of other processes result in
only secondary effects as can be seen in Figure~\ref{fig2} which displays
both the formation and destruction rates (multiplied by the neutral Li
fractional abundance $n({\rm Li})/n_{\rm H}$) of the dominant processes and
their time scales (per Hubble time $1/H(z)$). For $z>500$, the CMB field at
temperatures $T>1300$~K efficiently photoionizes any neutral Li atom
produced, but at lower redshifts the photoionization rate
(reaction~\ref{eq2}, L10) falls precipitously, allowing the rapid increase
in neutral Li through reaction~(\ref{eq1}, L5). Figure~\ref{fig2} shows
clearly that for $z<500$, the only important process is radiative
recombination~(\ref{eq1}, L5) with the radiative charge transfer reaction
\begin{equation}
{\rm Li} + {\rm H}^+ \rightarrow {\rm Li}^+ + {\rm H} + \nu
\label{eq3}
\end{equation}
\citep[L13,][]{sta96a} being only marginally significant with a
rate about two orders of magnitude smaller. Further, neutral Li
is mostly made over the narrow redshift interval $z\sim 500-300$,
before its formation time-scale becomes greater than the Hubble
time.

While the minor differences in the rate coefficients are unable to explain
the discrepancies between the Li and Li$^+$ abundances of GP98 and SLP98,
it seems likely that the adopted hydrogen recombination model may be
responsible for the difference. For $z<650$, GP98 obtained a smaller
electron fractional abundunce $x_{\rm e}$, with the SLD98 result being
$\sim 60\%$ larger by $z=100$. This trend is in agreement with the lithium
abundances displayed in Figure~\ref{fig1} indicating a significant
dependence on electron abundance, as would be expected. Given that the two
hydrogen recombination models are rather rudimentary, it would seem
appropriate to use an improved model. An ameliorated hydrogen recombination
calculation has been performed by \citet*{sea00} \citep[see also][]{sea99}
which allowed for the H excited state populations to depart from
equilibrium. Their calculation which consists of a 300-level model H atom
including all relevant bound-bound and bound-free transitions, finds a
change for the electron fraction of about 10\% from simple equilibrium
population results. \citet{sea99} have provided the program RECFAST which
emulates their detailed nonequilibrium population calculation. We have
coupled RECFAST to our code and obtained the lithium abundances shown in
Figure~\ref{fig1} where we have also continued to use the GP98 rate
coefficients for reactions~(\ref{eq1}) and (\ref{eq2}), and for the
remainder of this work. The neutral Li abundance has increased further as the
electron abundance computed by RECFAST is larger than obtained from both
GP98 and SLD98, being about 80\% larger than GP98 at $z=100$. Henceforth,
the electron abundances will be obtained from RECFAST.

The cosmological parameters (Model III of SLD96) used above have been
superseded by new values based on analysis of the latest CMB anisotropy
experiments \citep*[e.g.,][]{deb02,jaf01,wan01}.  To explore the range of
possible Li abundances allowable within the CMB uncertainties, we
considered the five models listed in Table~\ref{tab1}. Model A is
considered our fiducial case as all parameters are consistent with the best
fit values of \citet[the third column of their Table 2]{wan01}.  Model B
has the same values, but uses $h=0.75$, close to $h=0.72$
from the HST Hubble Key Project \citep{fre01}. 
Models A and B are also consistent with the
BOOMERANG experiment \citep{deb02}.  Models C and D follow the result of
\citet{jaf01}, $\Omega_{\rm b}h^2=0.032_{-0.004}^{+0.005}$, which is also
just barely outside of the 95\% confidence limit of \citet[see their Table
5]{wan01}. Model E is an extreme case taken as the upper limit from
\citet{jaf01}.  The corresponding primordial abundances ($Y$, $X_{\rm D}$,
and $X_{\rm Li}$) were obtained from \citet{bur01}.

\begin{figure}
\plotone{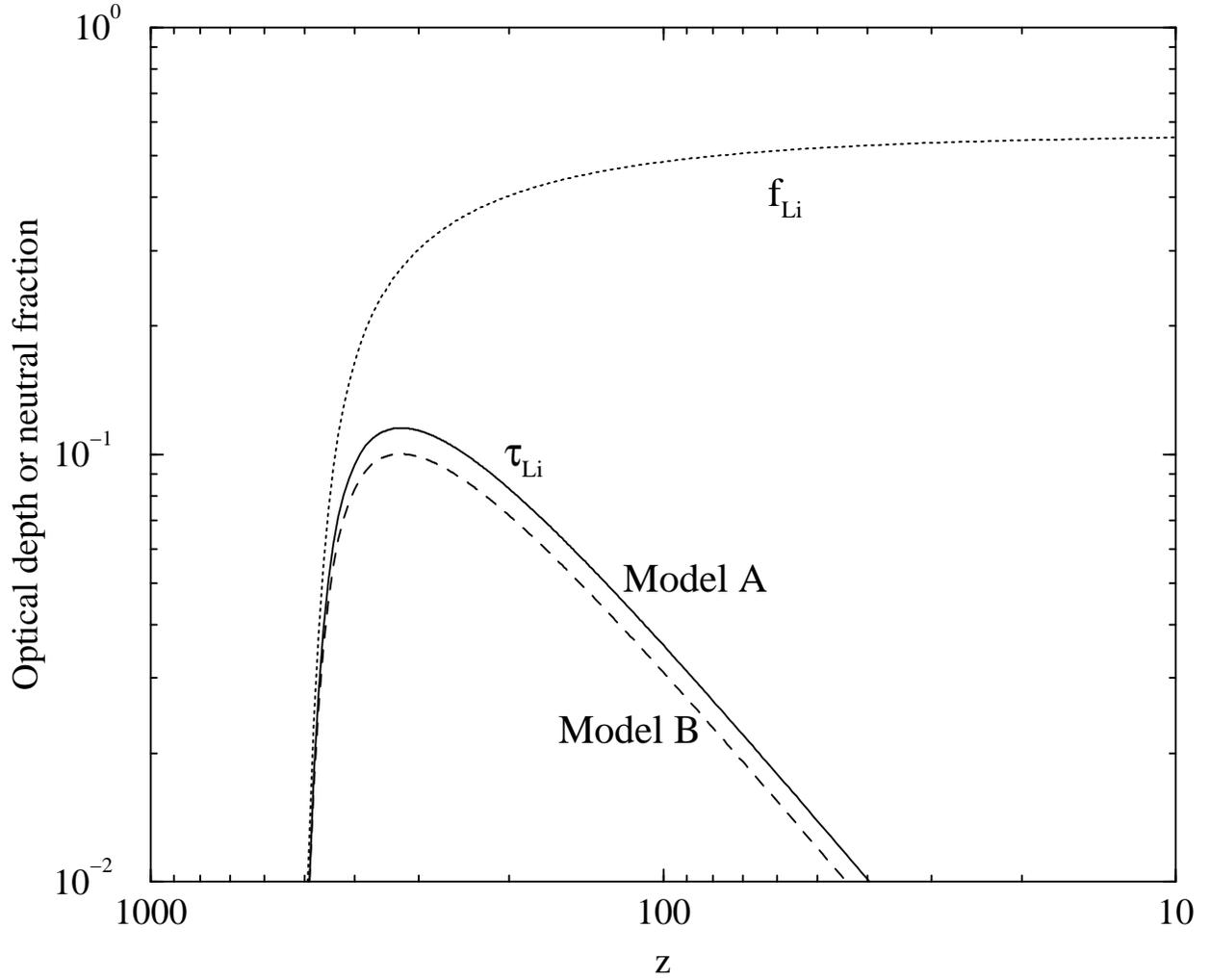}
\caption{The lithium neutral fraction and optical depth for Models A
($h=0.65$) and B ($h=0.75$). $\Omega_b h^2=0.02$ in both cases.
\label{fig3}}
\end{figure}

\begin{figure}
\plotone{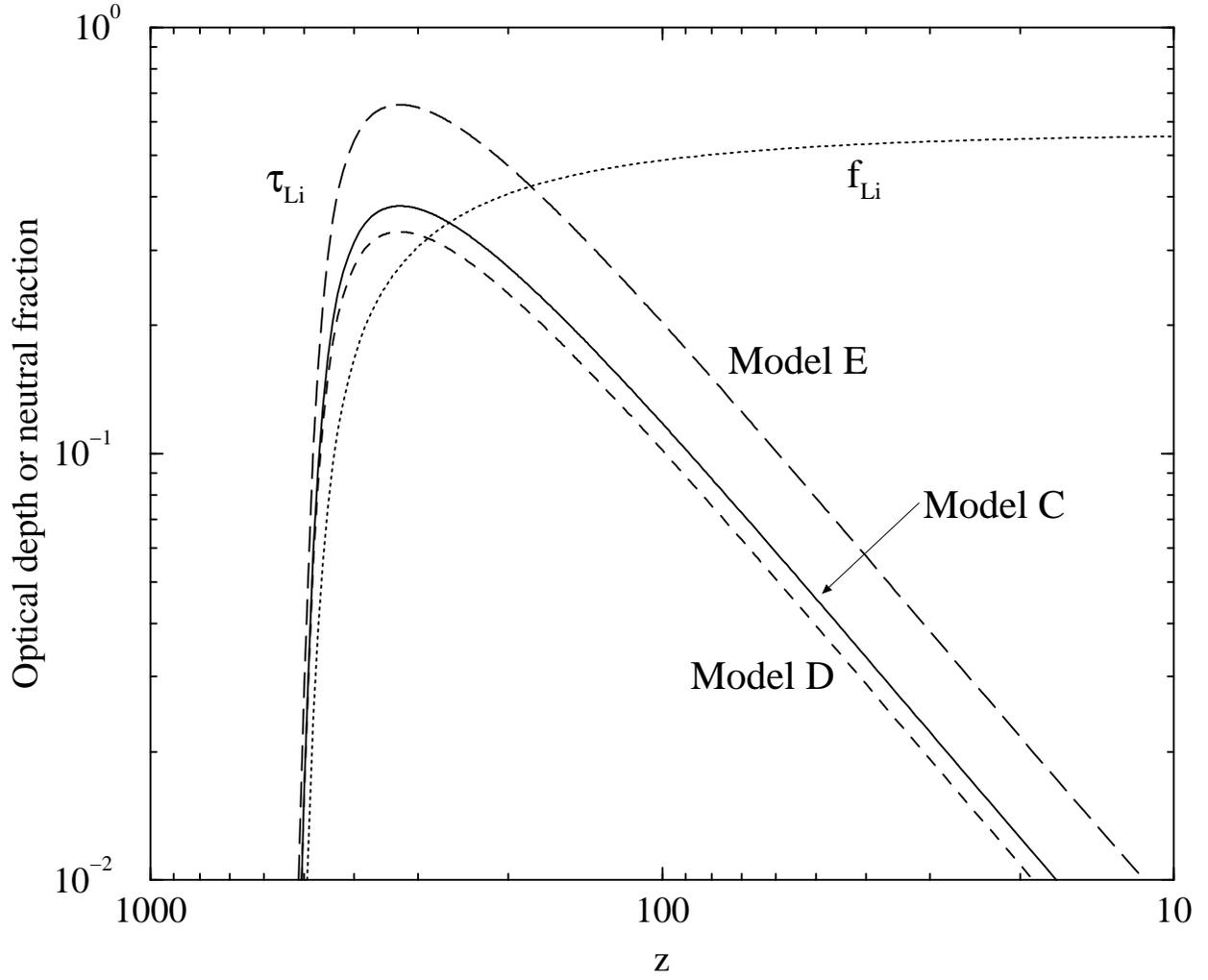}
\caption{The lithium neutral fraction and optical depth for Models C
($h=0.65$, $\Omega_b h^2=0.03$), D ($h=0.75$, $\Omega_b h^2=0.03$), and E 
($h=0.65$, $\Omega_b h^2=0.037$).
\label{fig4}}
\end{figure}

Using the electron abundances determined with RECFAST, the latest lithium
chemistry model of SLD98 with some improvements listed in \citet*{lep02},
but with the rate coefficients for reactions~(\ref{eq1}, L5) and
(\ref{eq2}, L10) replaced by the values from GP98, and the parameters listed
in Table~\ref{tab1}, we have calculated the new lithium neutral 
fractions\footnote{We do not consider the multiply charged ions Li$^{3+}$ and
Li$^{2+}$ as they were found by \citet{lep02} to have completely
converted to Li$^+$ by $z\sim 4000$ through sequential recombination.}
shown in Figures~\ref{fig3} and \ref{fig4}.  The neutral lithium fraction is
defined as
\begin{equation}
f_{\rm Li}(z)={{n({\rm Li})}\over{n_{\rm Li}}}
={{n({\rm Li})}\over{n({\rm Li}) + n({\rm Li^+})}},
\label{eq4}
\end{equation}
where
\begin{equation}
n({\rm Li}) = f_{\rm Li}(z) X_{\rm Li} n_{\rm H}(z)
\label{eq5}
\end{equation}
and
\begin{equation}
n_{\rm H}(z) = 1.123\times 10^{-5}(1-Y) 
\Omega_{\rm b} h^2 (1+z)^3~{\rm cm}^{-3}.
\label{eq6}
\end{equation}
The lithium neutral fractions $f_{\rm Li}$ are identical for 
Models A and B and for Models
C and D, while comparison of all the models reveal very little differences.
In fact, for all models $z_{\rm f}\sim 440$ and
$f_{\rm Li}=0.55-0.56$
(or $n$(Li$^+$)/$n_{\rm Li}= 0.44-0.45$) 
at $z=10$.

\section{The Optical Depth of Neutral Lithium}

The Sobolev optical depth of neutral lithium is given by
\begin{equation}
\tau_{\rm Li}(z)=
\sum_u {{\lambda_{1u}^3 A_{u1}}\over{8\pi}} {{g_u}\over{g_1}}
{{n({\rm Li})(1-b)}\over{H(z)}},
\label{eq7}
\end{equation}
\citep[see][equation A9]{sea00,bou97} where
\begin{equation}
H(z)^2 = H_0^2 \biggl [{{\Omega_0}\over{1+z_{\rm eq}}}(1+z)^4 +
\Omega_0(1+z)^3 + \Omega_{\rm K}(1+z)^2 + \Omega_{\Lambda}
\biggr ],
\label{eq8}
\end{equation}
$b$ accounts for the fraction of neutral Li in excited states, $g_u$ and
$g_1$ are the degeneracies of the upper and ground states, respectively,
and the sum is over the excited $2p~^2P_J$ fine-structure levels
$J=1/2,3/2$. We took $b=0$ and adopted the wavelengths 6707.76 and 6707.91
\AA~ for the $1/2\to 3/2$ and $1/2\to 1/2$ transitions from the
\citet{nis99} which also gives the transition probability
$A_{21}=3.72\times 10^7$ s$^{-1}$ for both transitions.  The optical depths
for Models A-E are plotted in Figures~\ref{fig3} and \ref{fig4}. In all cases,
$\tau_{\rm Li}$ is maximum near $z\sim 325$ where $f_{\rm Li}\sim 0.275$.
Over the five models, the maximum $\tau_{\rm Li}$ varies from $\sim 0.1$ to
$\sim 0.66$ which can be compared to the redshift independent optical
depths of 0.5--2 adopted by \citet{zal02}. Only the extreme Model E
approaches the optical depths considered by \citet{zal02}.  As expected
\citep[see][]{loe01,zal02}, the optical depth increases with $X_{\rm Li}$,
but the current results also demonstrate that $\tau_{\rm Li}$ increases
with $\Omega_{\rm b} h^2$ and decreases with $h$.

Making the approximation $H(z)=\Omega_0^{1/2}H_0(1+z)^{3/2}$ and
taking Model A as the fiducial case, we find that the optical depth reduces to
\begin{equation}
\tau_{\rm Li}(z)= 0.380 f_{\rm Li}(z) \biggl  
({{X_{\rm Li}}\over{3.907\times 10^{-10}}} \biggr ) 
\biggl ({{1-Y}\over{0.75286}} \biggr )
\biggl ({{\Omega_{\rm b}h^2}\over{0.02}} \biggr )
\biggl ({{0.65}\over{h}} \biggr ) \biggl ({{0.33}\over{\Omega_0}} \biggr )^{1/2}
\biggl ({{1+z}\over{300}} \biggr )^{3/2} ,
\label{eq9}
\end{equation}
which can be compared to equation (1) of \citet{zal02}.  The numerical
trends discussed above are evident in this relation in addition to
dependencies on $Y$ and $\Omega_0$. The shape of the optical depth curves
arise from a competition between the increasing neutral Li fraction $f_{\rm
Li}$ and the expansion of the Universe. As a consequence, the possible
effects of lithium on the CMB power spectrum are restricted to the epoch of
$z\sim 100-500$.

\section{Effects on CMB Anisotropies}

We have calculated the signatures of lithium resonant scattering on
the CMB anisotropies following the methods described in \citet{zal02}.
To isolate the effect of the scatterings we have kept all cosmological
parameters fixed as we varied the optical depth. As a reference model
we considered the standard LCDM model\footnote{For LCDM we chose
$\Omega_0=0.3$, $\Omega_{\Lambda}=0.7$, $h=0.7$ and $\Omega_bh^2
=0.02$.} and compared it with models having optical depths equal to
those of models A, C and E of the previous section.

\begin{figure}
\plotone{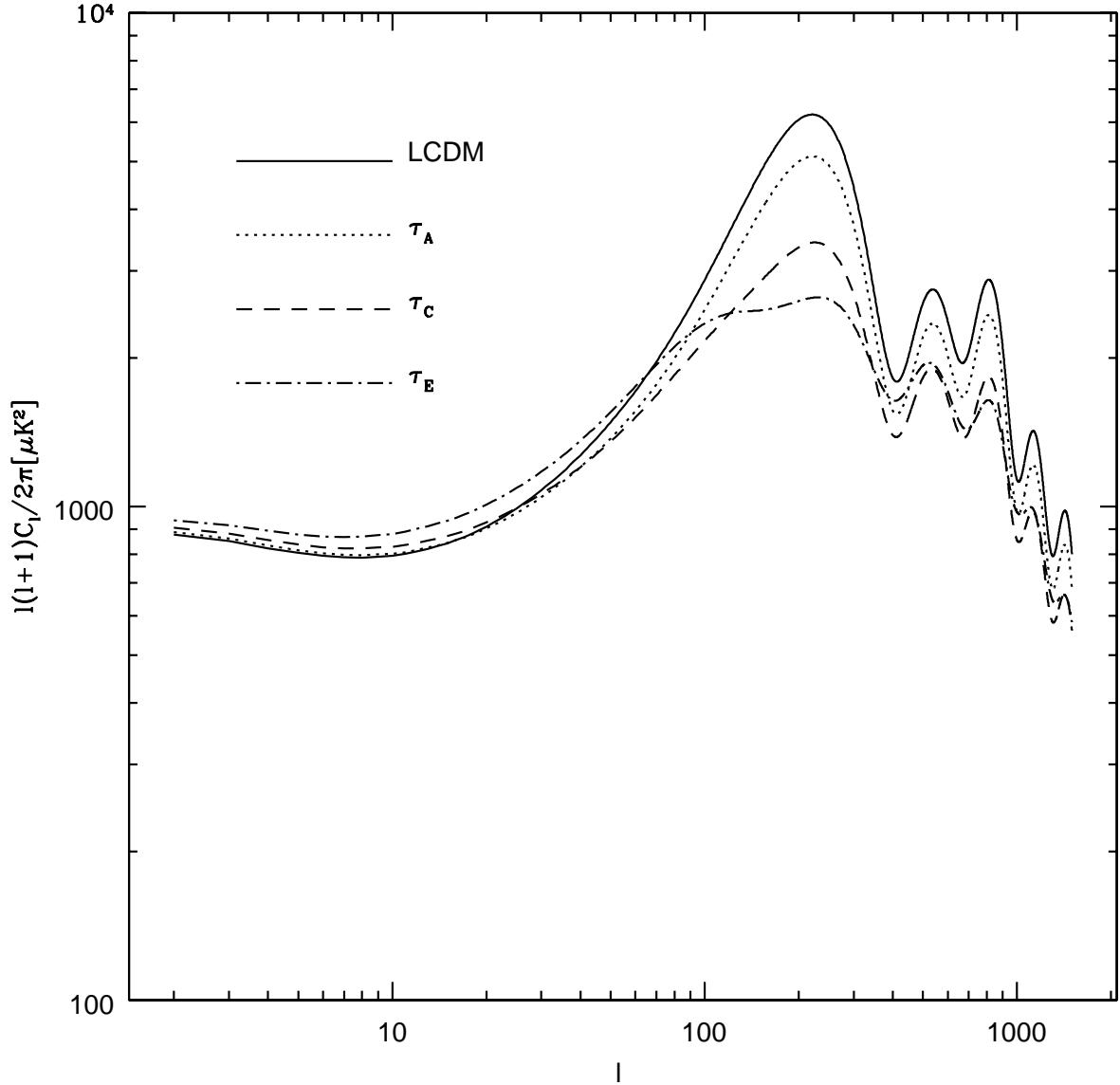}
\caption{Temperature power-spectra of the CMB anisotropies in models
with optical depths equal to those of models A, C, and E together the
spectra of the standard LCDM model without lithium scattering.  The
figure corresponds to an observed wavelength of $268.3~\mu$m.
\label{fig5}}
\end{figure}

\begin{figure}
\plotone{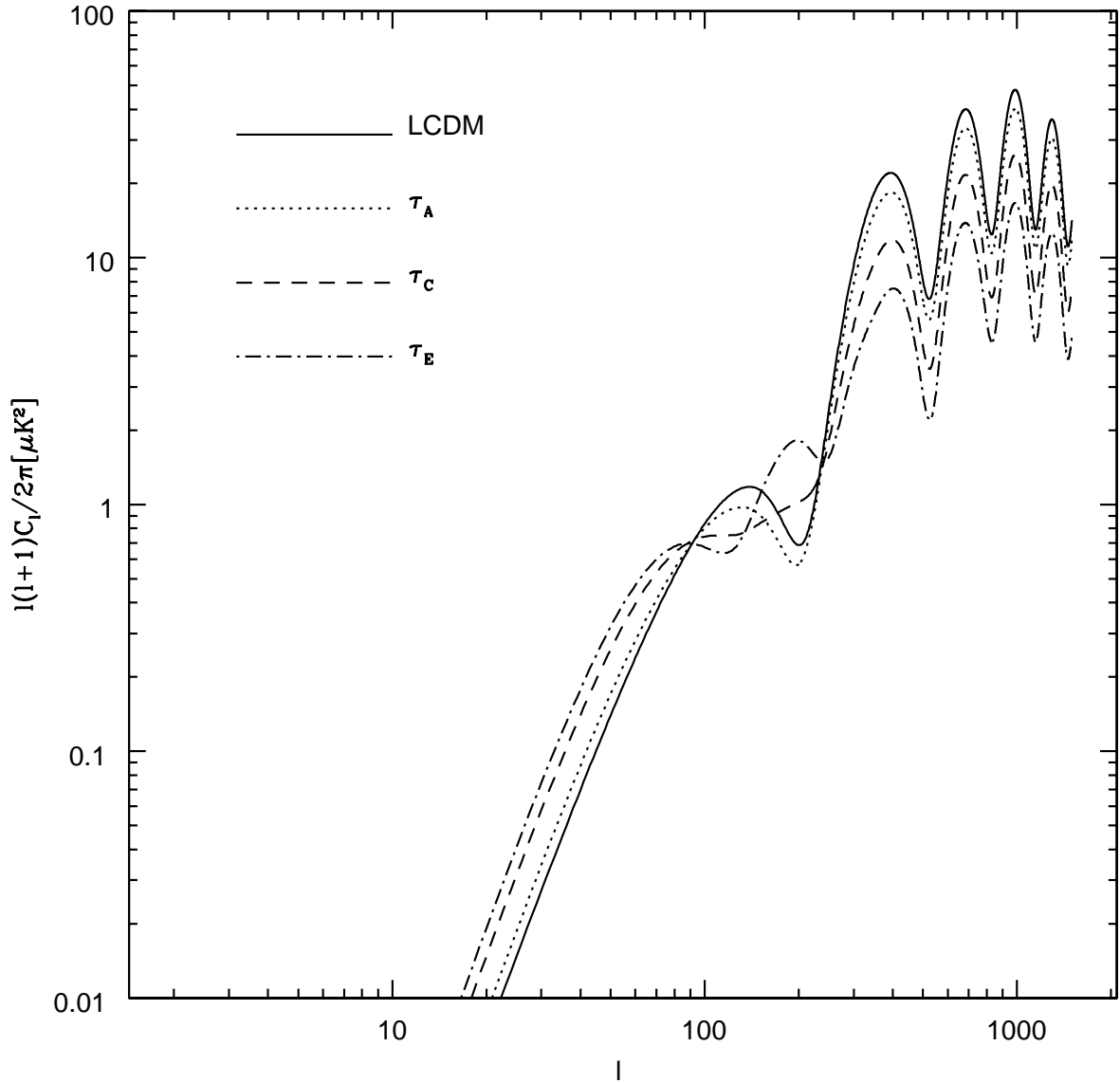}
\caption{Polarization power-spectra of the CMB anisotropies for models
with optical depths equal to those of models A, C, and E, at an
observed wavelength of $268.3~\mu$m compared to the standard LCDM
without lithium scattering.
\label{fig6}}
\end{figure}

In Figure \ref{fig5} we show the temperature anisotropy power spectra
predicted for a wavelength of $\lambda=268.3~\mu m$ (corresponding to
scattering at a redshift $z=400$). In Figure \ref{fig6} we show the
analogous plot for polarization. The effect of the scatterings
increases as the optical depth increases, so they are the smallest in
model A and the largest in model E.  The changes are dominated by two
contributions: (i) the Doppler anisotropies induced at the sharp
lithium scattering surface and (ii) the uniform $\exp(-\tau_{\rm
Li})$ suppression of the primary anisotropies which were generated at
decoupling.

At large multipoles $l$ (small angular scales), Figure \ref{fig6}
illustrates how the $\exp(-\tau_{\rm Li})$ suppression of the anisotropies
reduces the amplitude of power spectra by a factor $\exp(-2
\tau_{{\rm Li}})$. This effect leaves the shape of the power spectra
unchanged. For the models under consideration the optical depth ranges from
$0.1$ to $0.6$ so on these scales the power spectra are suppressed by a
factor between $0.3$ and $0.8$.

As was discussed by \citet{zal02}, the Doppler effect reaches a maximum on
degree scales and is responsible for the change of shape of the power
spectra on these scales. This effect can be noticed most dramatically in
model E for the temperature and in both models C and E for the
polarization.

\begin{figure}
\plotone{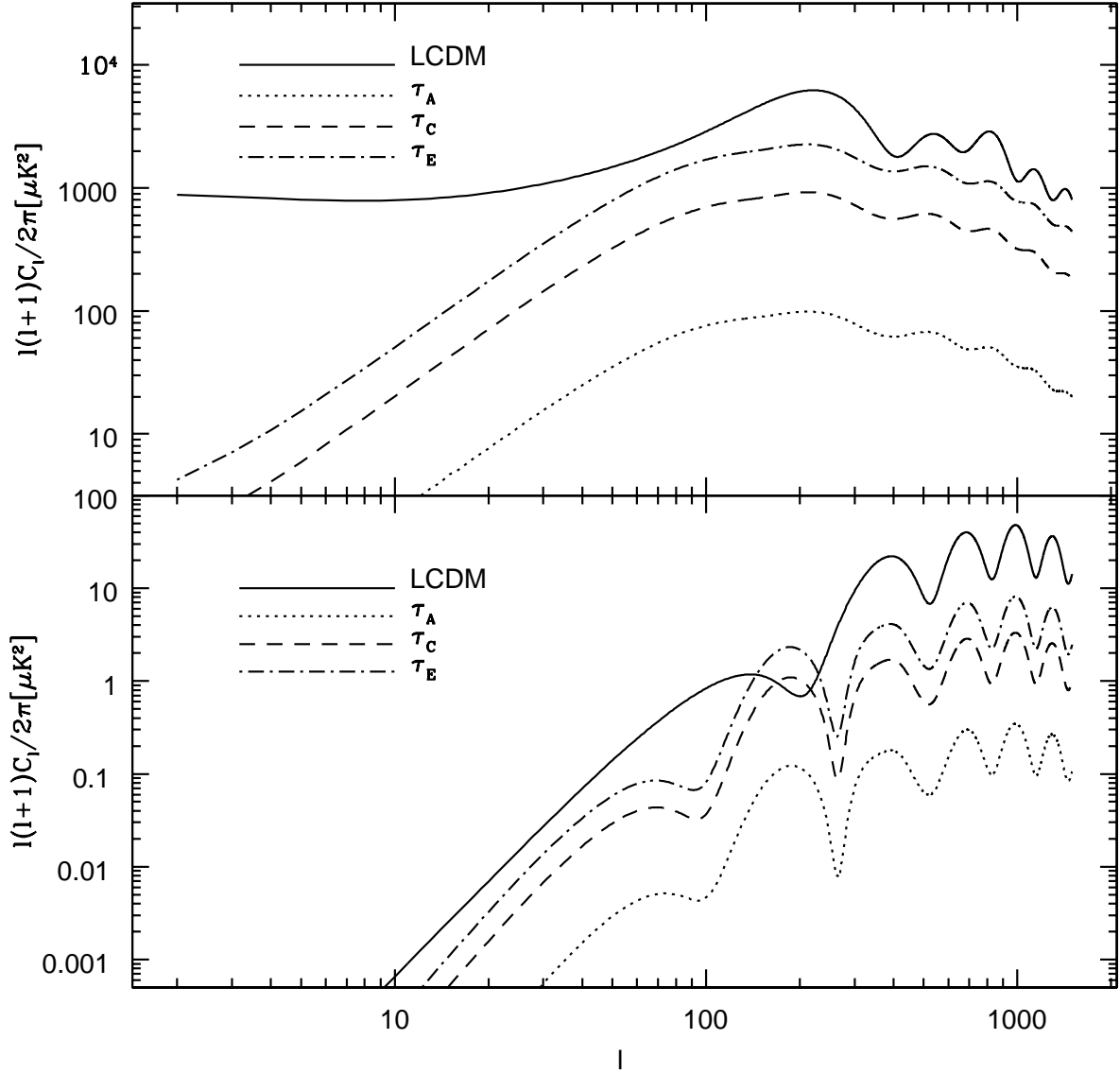}
\caption{Difference between a long wavelength power spectrum
(such as the one that will be produced by the {\it MAP} or {\it Planck}
satellites) and a power spectra at 268.3~$\mu$m when the optical depths are
those of models A, C, and E. We also show the spectra for the standard LCDM
model without lithium scattering. The top panel shows the results for
the temperature anisotropies and the bottom one for the polarization
anisotropies of the CMB.
\label{fig7}}
\end{figure}

Measurements of the CMB anisotropies at long photon wavelengths such as
those that will be made by the {\it MAP} and {\it Planck} 
satellites, will provide a
baseline for comparison against which the effect of lithium could be
extracted. These long wavelength maps, which are unaffected by the lithium
scattering, can be compared with maps obtained at shorter wavelengths.  In
particular a useful statistic to study is the power spectra of the
difference map (see Zaldarriaga \& Loeb 2001 for details). Here we subtract
the power spectra obtained at long wavelengths 
from that observed at 268.3~$\mu$m,
and consider the power spectrum of the residuals. The results for this
statistic are presented in Figure \ref{fig7}.

Figure \ref{fig7} indicates that the largest signal relative to the
unperturbed spectra is expected in polarization at degree angular scales
$(l\sim 200)$. In this range and for both models C and E, the power in the
difference spectrum is comparable to (or even larger than) 
the power in the long
wavelength spectrum.

\begin{table*}
\caption{Cosmological model parameters.}
\begin{tabular}{llllll}
\multicolumn{1}{l}{Parameter} & \multicolumn{1}{c}{Model A} &
\multicolumn{1}{c}{Model B} & \multicolumn{1}{c}{Model C} &
\multicolumn{1}{c}{Model D} & \multicolumn{1}{c}{Model E} \\
\tableline
$\Omega_0$       & 0.33     & 0.33  & 0.33   & 0.33  & 0.33 \\
$\Omega_{\rm K}$       & 0.0     & 0.0  & 0.0   & 0.0  & 0.0 \\
$\Omega_\Lambda$       & 0.67     & 0.67  & 0.67   & 0.67  & 0.67 \\
$z_{\rm eq}$     & 1.01e4 & 1.35e4 & 1.01e4 & 1.35e4 & 1.01e4 \\
$\Omega_{\rm b} h^2$ & 0.02 & 0.02 & 0.03 & 0.03 & 0.037 \\
$\eta_{10}$          & 5.479 & 5.479 & 8.219 & 8.219 & 10.14 \\
$\Omega_{\rm b}$ & 4.734e-2    & 3.556e-2   & 7.101e-2 & 5.333e-2 & 8.757e-2  \\
$h $             & 0.65     & 0.75   & 0.65   & 0.75  & 0.65 \\
$Y$                     & 0.24714   & 0.24714  & 0.25108   & 0.25108  & 0.25298 \\
$X_{\rm D}$   & 3.105e-5  & 3.105e-5 & 1.611e-5  & 1.611e-5 & 1.116e-5 \\
$X_{\rm Li}$  & 3.907e-10 & 3.907e-10 & 8.560e-10 & 8.560e-10 & 1.197e-9 \\
\end{tabular}
\label{tab1}
\end{table*}

\section{Discussion}

We have calculated the recombination history of lithium for a set of
different cosmological parameters (see Table 1). The resulting opacity to
the 6708~\AA~ resonant transition of neutral lithium (Figs. 3 and 4) distorts
both the temperature (Fig. 5) and polarization (Fig. 6) power-spectra of
the CMB anisotropies over a band of observed photon wavelengths around
$\sim 300~\mu$m.  The distortion is most pronounced in model E which
assumes a relatively high lithium abundance of $X_{\rm Li}=1.2 \times
10^{-9}$. Our detailed results agree with earlier estimates by Loeb (2001)
and Zaldarriaga \& Loeb (2001).  The predicted polarization signal is
comparable to the expected polarization anisotropies of the far-infrared
background (FIB; see Zaldarriaga \& Loeb 2001). A strategy for eliminating
the contribution from the brightest FIB sources would be helpful in
isolating the lithium signal.

The relevant wavelength range overlaps with the highest frequency channel
of the {\it Planck} mission ($352~\mu$m), with the Balloon--borne
Large--Aperture Sub--millimeter
Telescope\footnote{http://www.hep.upenn.edu/blast} (BLAST) that will have
$250,\ 350$ and $500~\mu$m channels and with the proposed balloon--borne
Explorer of Diffuse Galactic
Emissions\footnote{http://topweb.gsfc.nasa.gov} (EDGE) that will survey 1\%
of the sky in 10 wavelength bands between $230$--$2000~\mu$m with a
resolution ranging from $6^\prime$ to $14^\prime$ (see Table 1 in Knox et
al. 2001).  However, in order to optimize the detection of the lithium
signature on the CMB anisotropies, a new instrument design is required with
multiple narrow bands ($\Delta \lambda/\lambda\la 0.1$) at various
wavelengths in the range $\lambda=250$--$350~\mu$m. The experiment should
cover a sufficiently large area of the sky so as to determine reliably the
statistics of fluctuations on degree scales, and the detector should be
sensitive to polarization. For reference, the experiment should also
measure the anisotropies at shorter wavelengths where the FIB dominates.
All of these requirements, of course, will make the detection of the
lithium signature a difficult experiment. The difficulty cannot be fully
accessed until more is known about the FIB and the sources responsible for
it. Some considerations and strategies are discussed in
\citet{zal02}. Further, in order to detect the effect of lithium, high
signal-to-noise maps of the primordial CMB at long wavelengths should be
made for the same region of the sky.  Fortunately, these maps will become
available from future CMB missions such as the {\it Planck} satellite.

On the theoretical front, improvements on the predicted lithium
recombination history can be made by performing an explicit calculation of
the non-equilibrium populations of the excited states in the same
level-by-level fashion as that of the hydrogen recombination calculation of
\citet{sea99,sea00}.  For hydrogen it was found that the excited state
levels were overpopulated for $z < 800$ as the levels fall-out of
equilibrium with the background radiation field. A slower photoionization
rate coupled with faster cascade rates results in a faster net recombination
rate ultimately resulting in a $\sim$10\% change in the ionization
fraction. \citet{sea99,sea00} also confirmed that all Lyman transitions are
optically thick during hydrogen recombination indicating that Case B
recombination holds at $z > 800$ but not at lower redshifts.  The optical
depths computed here for lithium are less than one for the ground state
transition suggesting that all other transitions should be optically thin,
indicating Case A recombination. The Case A lithium equilibrium
recombination rate coefficient of VF96 might result in an
underestimation of the neutral lithium fraction (and the optical depth),
but probably only by $\sim 10\%$.

We also note that non-conventional recombination histories may have a
substantial impact on our predicted signal. For example, it has 
recently been proposed that recombination could have been delayed by
the presence of additional radiation at $z\sim 10^3$, possibly  
from stars, AGNs, or accretion by primordial
compact objects \citep*{pee00,mil01}. The neutral fraction
of lithium (and its optical depth) could be larger than we calculated based
on the standard recombination history, as the neutral lithium fraction is
sensitive to the residual electron abundance. However, the stronger
radiation field might override any gains in recombination of lithium due to
the increase in its photoionization rate. Models with detailed radiation
fields are needed to calculate the net effect that these processes
have on the lithium optical depth.

Finally, while experiments to observe the lithium distortion on the
temperature and polarization CMB power spectra appear to require 
redesign of detectors and new observational strategies, the benefits
as mentioned in \citet{zal02} could potentially be significiant:
\begin{enumerate}
\item If the lithium signal could be separated from the FIB contamination,
it would offer the possibility of constraining the
lithium primordial abundance. Inferring the lithium primordial
abundance from stellar observations is known to be complicated
by stellar lithium depletion and galactic lithium production.
Further, the lithium primordial abundance is a sensitive
indicator of the baryon abundance, its mean value as well as
inhomogenities. It would then provide a possible means to discriminate
among Big--Bang nucleosynthesis models.
\item The lithium signature on the CMB anisotropies is the only probe
proposed so far for structure in the dark ages of the early Universe.
Other methods, such as the suggestion by \citet{ili02} that angular
fluctuations in 21 cm emission from minihalos could probe redshifts between
reionization and $z\sim 20$, rely on the existence of collapsed objects and
do not reach the high redshifts ($z\sim$100 to $\la 500$) probed by the
lithium signal (see review by Barkana \& Loeb 2001).  The lithium and 21 cm
signals would both give measures of the baryonic density fluctuations, the
latter during the era of early star formation leading to reionization and
the former just before the condensation of the first objects.
\end{enumerate}
\acknowledgments

This work was supported in part by NASA grants NAG 5-7039, 5-7768, and
by NSF grants AST-0071019 (AL), AST-0087172 (PCS), AST-0087348 (SL),
AST0088213 (AD) and AST-0098506, PHY-0116590 and the David and Lucille
Packard Foundation (MZ). We thank Sara Seager and Gary Ferland for
helpful discussions and Daniel Galli for supplying his numerical
results. PCS thanks the Institute for Theoretical
Atomic and Molecular Physics at the Harvard-Smithsonian Center for
Astrophysics, funded by the NSF, for travel support.

\end{document}